# Facilitating Evolution during Design and Implementation

Richard McClatchey

*Centre for Complex Cooperative Systems, UWE, Bristol BS16 1QY, UK*
Tel: +44 (0)117 3283176  Fax: +44 (0)117 3283134  Email: Richard.McClatchey@uwe.ac.uk

Abstract: The volumes and complexity of data that companies need to handle are increasing at an accelerating rate. In order to compete effectively and ensure their commercial sustainability, it is becoming crucial for them to achieve robust traceability in both their data and the evolving designs of their systems. This is addressed by the CRISTAL software which was originally developed at CERN by UWE, Bristol, for one of the particle detectors at the Large Hadron Collider, which has been subsequently transferred into the commercial world. Companies have been able to demonstrate increased agility, generate additional revenue, and improve the efficiency and cost-effectiveness with which they develop and implement systems in various areas, including business process management (BPM), healthcare and accounting applications. CRISTAL's ability to manage data and its semantic provenance at the terabyte scale, with full traceability over extended timescales, together with its description-driven approach, has provided the flexible adaptability required to future proof dynamically evolving software for these businesses.

## Background to the CRISTAL System

Organisations of many types increasingly operate within environments that dictate unforeseeable changes. Systems need to be able to evolve dynamically in response to changes in technology and consequently there must be full traceability between the design and the evolving system specification. Research at UWE's Centre for Complex Cooperative Systems (CCCS) from 1997 onwards (see [1] and [2]), has addressed the need to be able to develop software whose specifications evolve beyond its design phase and may have no definite endpoint. CCCS has developed an approach that allows systems to reconfigure themselves dynamically, enabling software to be versioned and rolled out into production to sit seamlessly alongside existing live systems without designer intervention. Following such a semantic-based approach it has been shown [2] that flexibility, interoperability, adaptability and simplicity of maintenance can be optimized. This approach has been applied, in collaboration with CERN (Switzerland) and CNRS (France), to the creation of a new software development environment called CRISTAL for CERN's CMS (Compact Muon Solenoid) experiment [3], to address that experiment's software needs over its extended design timeline.

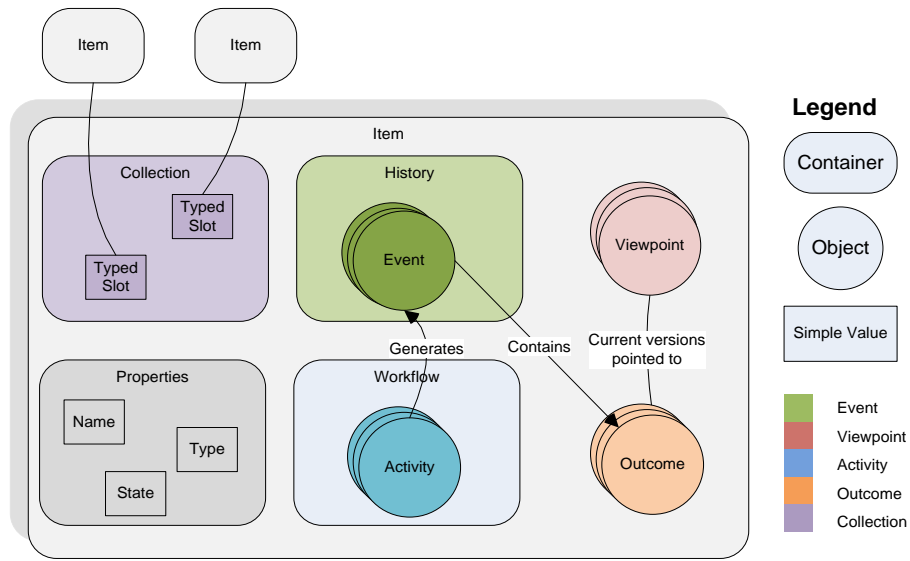

Figure 1. The components of adescription-driven Item in CRISTAL

CRISTAL embodies a "description-driven" approach [2]: all logic and data structures are described by metadata, whose semantics are captured in a model which can be modified and versioned online as the design of the system (in this case the experiment's particle detector) changes (see figure 1). Our research has shown that the creation of a flexible system can be facilitated by modelling the metadata: the resulting software is reusable across applications and can handle complexity, version control and system evolution. CRISTAL was used to calibrate CMS and helped in its apparent discovery of the Higgs Boson in 2012 (figure 2). The advantages of separating design, implementation and instantiation further vindicate the use of such "meta-models" (see [4] and [5]).

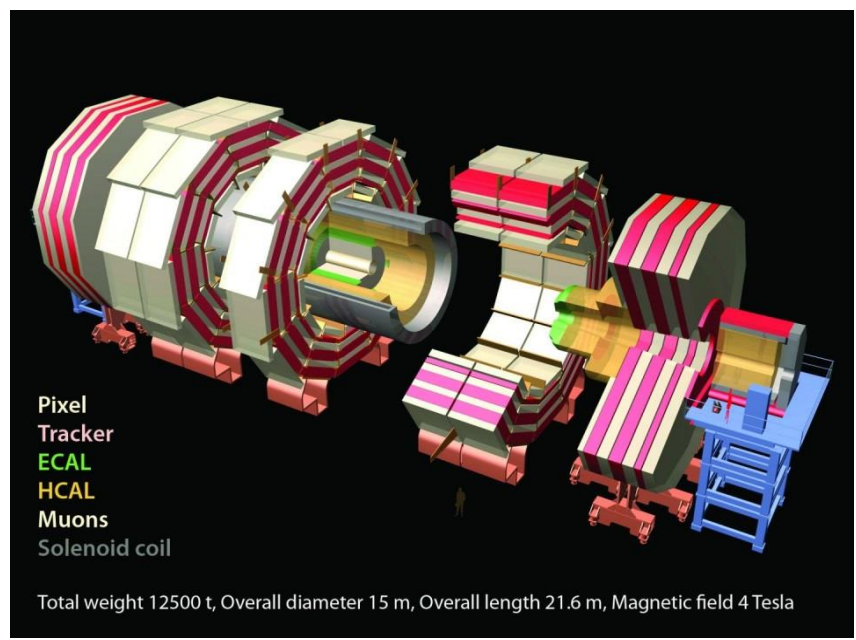

Fig 2: CMS at CERN part of whose construction was managed by CRISTAL (CERN Photo).

During the construction period of the CMS Electromagnetic Calorimeter (ECAL [3]) from 1999 to 2008, over 70,000 individual lead tungstate crystals were characterised (each yielding 3-5 Mbytes of data) and their data provenance captured in CRISTAL with only minimal interruptions to round-the-clock operation. CRISTAL only needed to be upgraded seven times during this period of near-continuous operation, and of those only one was a required update that effected end-users. The underlying semantic model used to store CMS construction data evolved many times between 2000 and 2008 but CRISTAL provided uninterrupted data taking for the scientists and ease of system maintenance. After the production-oriented workflows of CMS, CRISTAL was targeted at capturing and storing scientific data provenance and workflow orchestration [6] in the EC-funded neuGRID project (2008-2011) and its follow up neuGRIDforUsers project (N4U, 2012-2015). Clinical researchers in Italy, Netherlands and Sweden have been using CRISTAL on the neuGRID infrastructure to support ongoing studies of Alzheimer's Disease biomarkers using CRISTAL to orchestrate their analysis procedures and to log outcomes as part of their "Virtual Laboratory" (shown in Figure 3). The richness of the CRISTAL model in terms of its knowledge representation enables the integration of heuristic-based searches and the potential for the use of machine learning techniques for the optimization of analysis procedures in clinical research.

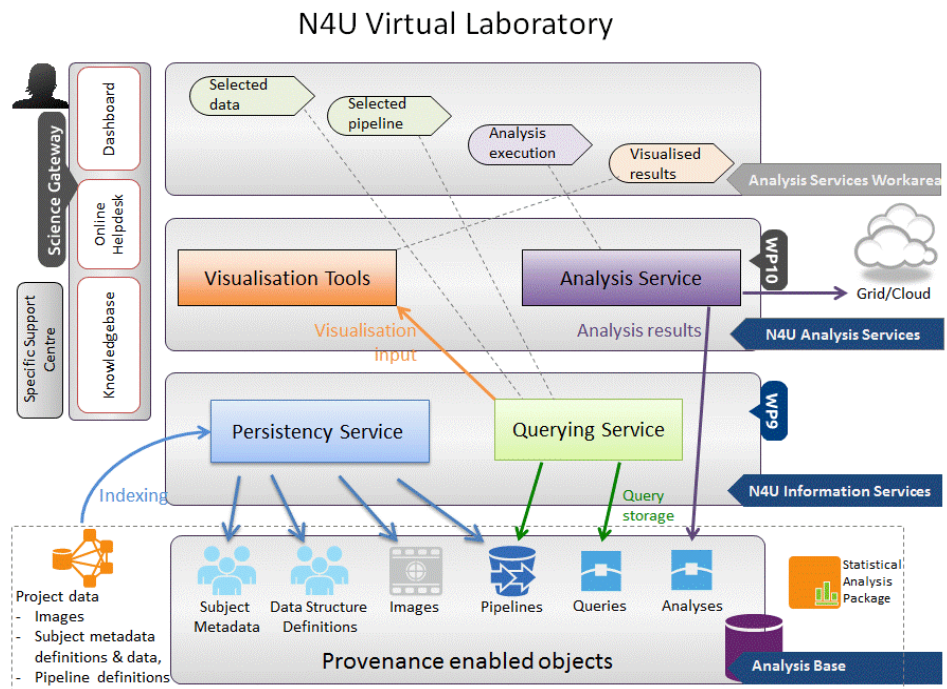

Fig 3: The N4U Virtual Laboratory with CRISTAL supported provenance.

# Transfer to Industry

The development of CRISTAL for managing workflows and product-related data was conducted in close consultation with user communities – initially physicists and engineers at CERN and latterly health practitioners. CRISTAL was made available by UWE to the CMS ECAL user community from 2003; it also became the basis for a programme of external exploitation by M1i, France, in Business Process Management (BPM). At M1i it has demonstrated its ability to be responsive to changing user requirements and to support on-the-fly system evolution over extended product lifecycles, due to the flexibility of its underlying model. This has enabled M1i to supply its customers with guaranteed system response and the ability to cater for changing requirements as well as flexibility in data management and simplicity in system support. CRISTAL can handle the complexity of data-intensive systems and provide the flexibility to adapt to the changing scenarios typical of any production management system, or indeed any process in which workflow and data traceability is crucial, including those employing AI solutions.

Via the product name Agilium it has been sold by M1i into the retail, finance and manufacturing sectors of European industry, for use in the area of Business Process Management (BPM) (see figure 3). Agilium used the CRISTAL kernel for workflow and process traceability, and also for the integration and co-operation of multiple business processes – especially in business-to-business applications. The M1i product enables business processes to be harmonised and orchestrated in operation using a CRISTAL database. It tracks their workflows and integrates multiple potentially heterogeneous processes, such as order processing, stock control, sales management and business logistics thus providing enterprise-wide data and information management. This has found application in real-world industrial contexts, for example the systems used by Nexcis for photovoltaic cell production, by the Ville de Lyon (France) for managing its purchasing /ordering and operational procedures, and by the STTS specialist aerospace painting/sealing for managing its internal business process controls. In particular Agilium enables these customers to trace their data across applications and to adapt to system evolution with little or no disruption to their live systems.

CRISTAL-Agilium also integrates the management of data coming from different sources and unites Business Process Management with Business Activity Management (BAM) and Enterprise Application Integration (EAI) through the capture and management of their designs in the CRISTAL database. This has been applied at the Bayer Group, where CRISTAL-Agilium has enabled Bayer customers to conduct and handle business-to-business transactions and to facilitate the management of targeted application domains (e.g. logistics, retail and government domains). Other CRISTAL-Agilium clients of M1i include Dynastar, GDP Vendome, the ski resorts of Tignes and Meribel, Photowatt

Techniologies, SoTRADEL and Midor. The software is used to manage the business processes of and between these commercial companies enabling M1i to gain a unique position in the BPM market (as recognised by the Gartner Group, 2009). Income generated by CRISTAL-Agilium licence sales by M1i in 2008-2013 top €1.6M.

## Further Exploitation

CRISTAL has been recently further exploited by the Technoledge start-up company in Switzerland. Starting early 2012, Technoledge has been working closely with UWE to adapt CRISTAL to domains outside of BPM. Amongst other things, application of CRISTAL by Technoledge has demonstrated, for the first time, full traceability from raw materials to final product facilitated by the use of CRISTAL in manufacturing execution systems. This has led to the following outcomes:

- Building on the way CRISTAL was used for experiment at CERN to integrate different data structure versions together, Technoledge have deployed it to provide the support environment for the development of production lines to manufacture next-generation sustainable fuel cells for future electric vehicles.
- Technoledge has exploited CRISTAL to integrate previously outsourced software packages for a major French accountancy firm, taking advantage of CRISTAL's ability to manage several contrasting models in the same workspace.
- Since 2008 CRISTAL has been used by the neuGRIDforUsers (N4U) project to provide a system for clinicians investigating complex 3D-imaging techniques to study cortical thickness as a biomarker for Alzheimer's disease. They have used CRISTAL to track the analysis of complex algorithms and large data sets to help identify patients who may be susceptible to mild cognitive impairment that could lead to the onset of dementia. Early identification of such conditions is enabling doctors in medical centres across Europe to diagnose Alzheimer's Disease and thereby to prescribe suitable drug therapies to slow its onset. CRISTAL plays an essential role in providing a 'Virtual Laboratory' as shown in figure 3 for supporting and logging the ongoing developments in Alzhemier's Disease research (see [6] and [7]). CRISTAL has enabled them to systematically manage their semantic data and processes in a maintainable, more flexible and evolvable, and thus profitable, manner.

CRISTAL is being released in Q3 of 2014 under a LGPL V3.0 Open Source licence. The software is available from the CRISTAL-ISE project web pages at www.cristal-ise.org or via contact with the author of this paper. The software will be supported by an Open Source community of users who are free to develop its Kernel or application 'modules' that use the Kernel for any domain. It is planned that this community of users will openly share developments under the LGPL licencing arrangements.

# Futures

CRISTAL-iSE is an ongoing (2012-2016) project about fostering collaboration between industrial partners and academia and the development of individual researchers in the project. The partners involved in the project are from UWE, M1i (Annecy, France) and Alpha-3i (Rumilly, France) [8]. At the end of the project it is foreseen that there will be three final pieces of software developed. These will be a new open source version of CRISTAL, a new version of Agilium based on this version of CRISTAL and Cimag-RA (Alpha3i) which will be built upon both the work developed by UWE and M1i. In the Cimag-RA application, both the new versions of Aglium and CRISTAL will be used to create a Resource Allocation application. From the large datasets that are available already, a PROV-compliant provenance model will be created to foster collaboration between the wider provenance research communities in order to maximise the commercial impact of the CRISTAL developments.

Acknowledgements : The development of CRISTAL has been made possible by the support of CERN, CNRS and UWE and colleagues therefrom and in the context of projects supported by the European Commission.

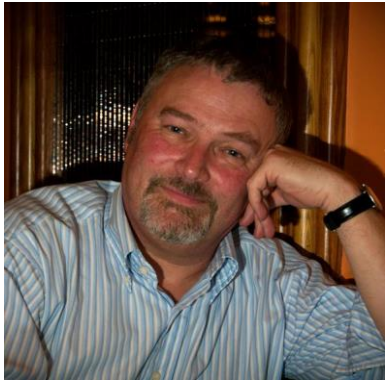
**Prof. McClatchey** has been research active for the past 30 years and has led many projects funded by industry and by the EC in the areas of large-scale distributed data and knowledge management, in data modelling and in systems design and integration. Currently a Fellow of both the British Computer Society and the Institute of Engineering and Technology with a PhD in Physics (Sheffield, 1982) and DPhil in Computer Science (West of England, UWE 1999), McClatchey has published over 200 papers and has held the Chair of Applied Computer Science at UWE since2000. His current research interest lies in Cloud data and knowledge representation and management and particularly in its application in medicine and smart cities. He currently leads the Centre for Complex

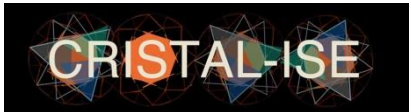 Cooperative Systems at UWE and is active in collaborative projects at CERN, and with many international partners in numerous EC projects including Health-e-Child, SHARE, neuGRID/N4U and CRISTAL-ISE. He has chaired several international conferences, workshops and symposia during his 23 years at UWE, Bristol.